\documentclass[conference]{IEEEtran}
\IEEEoverridecommandlockouts
\usepackage{svg}
\usepackage{amsmath}
\usepackage{amsfonts}
\usepackage{amssymb}
\usepackage{graphicx}
\usepackage{epstopdf}%
\usepackage[nice]{nicefrac}
\usepackage{algorithmic}
\usepackage{algorithm}
\usepackage{array}
\usepackage{authblk}
\usepackage{tikz}
\usepackage[utf8]{inputenc}
\usepackage{pgfplots} 
\usepackage{pgfgantt}
\usepackage{pdflscape}
\usepackage{pdfpages}%%%%%%%%%%%%%%%%%%%%%%%%%%%
\pgfplotsset{compat=newest} 
\pgfplotsset{plot coordinates/math parser=false}
\usepackage{grffile}
\usetikzlibrary{plotmarks}
\usepackage{amsthm}
\usepackage{amsmath}%
\usepackage{MnSymbol}%
\usepackage{wasysym}
\usepackage{subcaption}
\usepackage[mathscr]{euscript}
\usepackage{cite}
\usepackage{amsfonts}
\usepackage{pgf}
\usepackage{mathtools}
\usepackage{graphicx}
\usepackage{algorithmic}
\usepackage{algorithm}
\usepackage{array}
\usepackage{tikz}
\usepackage[utf8]{inputenc}
\usepackage{pgfplots} 
\usepackage{pgfgantt}
\usepackage{pdflscape}
\pgfplotsset{compat=newest} 
\pgfplotsset{plot coordinates/math parser=false}
\usepackage{grffile}
\usetikzlibrary{plotmarks}
\usepackage{tikzscale}
\usepackage{textcomp}
\usepackage{algorithmic}
\usepackage{algorithm}
\usepackage{array}
\usepackage{tikz}
\usepackage[utf8]{inputenc}
\usepackage{pgfplots} 
\usepackage{pgfgantt}
\usepackage{pdflscape}
\pgfplotsset{compat=newest} 
\pgfplotsset{plot coordinates/math parser=false}
\usepackage{grffile}
\usetikzlibrary{plotmarks}
\usepackage[T1]{fontenc}
\usepackage{graphicx}
\usepackage{overpic}
\usepackage{array}
\usepackage{color}
\usepackage{url}
\usepackage{tikz}
\usepackage[utf8]{inputenc}
\usepackage{pgfplots} 
\usepackage{cite}

\usepackage{amsmath, amsthm, amssymb}

\def\CN{\mathcal{C}\mathcal{N}} %Complex Gaussian
\usepackage{amsmath}
\usepackage{algorithmic}
\begin{document}
\title{A Nonlinear Autoregressive Neural Network for Interference Prediction and Resource Allocation in URLLC Scenarios}


\author{Christian Padilla}
\author{Ramin Hashemi}
\author{Nurul Huda Mahmood}
\author{Matti Latva-aho}
\affil{Centre for Wireless Communications (CWC), University of Oulu, Oulu, Finland, \authorcr Emails: {\{christian.padillaolivo, ramin.hashemi, nurulhuda.mahmood, 	matti.latva-aho\}@oulu.fi}
\thanks{This work was supported by the Academy of Finland 6Genesis Flagship project under grant 318927.}}

\maketitle

\begin{abstract}
Ultra reliable low latency communications (URLLC) is a new service class introduced in 5G which is characterized by strict reliability ($1-10^{-5}$) and low latency requirements ($1$ ms). To meet these requisites, several strategies like over-provisioning of resources and channel-predictive algorithms have been developed. This paper describes the application of a Nonlinear Autoregressive Neural Network (NARNN) as a novel approach to forecast interference levels in a wireless system for the purpose of efficient resource allocation. Accurate interference forecasts also grant the possibility of meeting specific outage probability requirements in URLLC scenarios. Performance of this proposal is evaluated upon the basis of NARNN predictions accuracy and system resource usage. Our proposed approach achieved a promising mean absolute percentage error of 7.8 \% on interference predictions and also reduced the resource usage in up to 15 \% when compared to a recently proposed interference prediction algorithm. 

%As a performance reference, an algorithm developed by Nurul [reference here] will be used. 

\end{abstract}


% \fi
\IEEEpeerreviewmaketitle

\section{Introduction}
Ultra-reliable low-latency communication (URLLC) is a service class introduced in 5G that targets strict quality of service (QoS) requirements: reliability of $1-10^{-5}$ and latency of 1 ms \cite{Sachs}. These strict QoS guarantees can be achieved by over-provisioning of resources, i.e., allocating a large amount of resources in order to guarantee low outages \cite{mahmood2019multichannel}. However, such an approach is neither resource efficient nor scalable. Therefore it is necessary to develop a radio resource management (RRM) technique that overcomes these scalability and resource usage (RU) challenges while enabling URLLC \cite{Nurul}. 

Interference management has always been a challenge in wireless systems design~\cite{schmidt2021interference}. Efficient interference management is among the possible ways to ensure efficient resource allocation (and to ensure scalability) for URLLC services. Having the ability to predict future interference is beneficial to tackle the effects of interference \cite{Zheng_2014}, resources can be allocated efficiently using this knowledge. Obtaining accurate forecasts of the interference power is a challenging task due to the time varying nature of wireless channels. 
Various researchers have proposed different interference prediction techniques. 
Haenggi et al. \cite{5934671} used stochastic geometry to derive mean interference and probability distribution in wireless networks, Schmidt et al. \cite{schmidt2021interference} proposed the use of a recursive predictor to estimate future interference values by filtering the measured interference. Varma presented a two-staged machine learning algorithm to remove multiple interference from wireless signals while minimizing information loss in \cite{varma}.
% Chinchali et al. \cite{Chinchali_deeplearning} obtained promising results segmenting corrupted wireless transmissions into desired signal and interference estimate using Deep Learning.

%we propose an AI-based prediction scheme to forecast the interference power in a wireless system
%since the resources needed will be calculated using an interference value that is proportional to the true value in order to meet URLLC strict requirements.
%URLLC enables mission critical communications, like reliable remote
% action with robots or coordination among vehicles \cite{popovski}.

% Predicting the interference levels . 
Methodologically speaking, the interference power prediction can be classified into statistical and artificial intelligence (AI) methods. Statistical interference power prediction implies that some statistical properties of the interference power, such as mean interference, or even full probability distributions, are used to estimate future interference values. AI-based interference prediction consists of using a machine learning algorithm to approximate the behavior of the channel with a model that uses past interference values as inputs and future interference values as outputs. In this proposal we consider an AI-based solution since there is a broad study of statistics-based interference prediction techniques but there is scarcity in the application of AI for the same purpose.  

A Nonlinear Autoregressive Neural Network (NARNN) has been chosen for interference prediction due to its high forecasting accuracy \cite{wind} and the fact that there is no need of a mathematical model of the process other than its own universal algorithm for future series prediction. A major benefit of NARNNs is that they accept dynamic inputs represented by time series sets where past values are used to predict future values in a time series set \cite{en9090684}. Several researchers have demonstrated promising prediction capabilities of NARNNs. Gairaa et al. \cite{solar} used a NARNN to predict global solar radiation levels, Ruiz et al. \cite{en9090684} used a NARNN to forecast energy consumption in public buildings and Sarkar et al. \cite{wind} studied the performance of NARNNs for long-term wind speed forecasting. 
% For this proposal, the NARNN training data sets consist of interference power time series obtained from computer simulations of a downlink with N-1 interferers and a Rayleigh channel.  To evaluate the performance of the proposed NARNN, MAPE and MSE indicators have been used. Resources are allocated depending on NARNN predictions and specific target outage probabilities.

The layout of this paper is as follows: the system model is presented in Section~\ref{sec:systemModel}. The NARNN design is described in Section ~\ref{sec:narModel}. In Section~\ref{sec:results}, the results are discussed. The conclusion of this proposal is presented in Section ~\ref{sec:conclusion}.
\section{System Model}
\label{sec:systemModel}
We consider the downlink of a typical multi-user wireless system in the two dimensional space organized by $N$ transmitters (i.e., base stations (BS)), each of which serves a single receiver. We focus our attention on a particular receiver of choice (denoted as user $n$), which is receiving a URLLC transmission in the presence of interference from the remaining $N - 1$ BSs. We assume that the transmitting BS of interest is the closest BS, while the remaining $N - 1$ interfering BSs are distributed across the service area at varying distances from the receiver of interest. 

The corresponding channel responses between a transmitter and a receiver follows circularly symmetric complex Gaussian distribution. It is worth noting that the adaption of the channels' distribution can be generalized to more practical and complex scenarios without loss of generality. In order to keep the latency budget as low as possible, we assume a single-shot transmission scheme. Therefore, the received signal at the receiver is given by
\begin{flalign}
    r = h_nx_n + \underset{\text{Interference}}{\underbrace{\sum_{k=1,k\neq n}^{N}h_kx_k}}+w,
\end{flalign}
where $h_k\sim \CN(0,\beta_k)$, $\forall k \in \{1, \ldots, N\}$ denotes the channel response between the $k^{th}$ BS and the user $n$ with path loss $\beta_k$. Also, the white Gaussian noise $w$ has variance $\mathbb{E}[|w|^2]=\sigma^2$ and the symbols transmitted from BS to each user $x_n$ where $\mathbb{E}[|x_n|^2]=p_n$ denotes the maximum transmit power. Thus, the signal-to-interference-plus-noise ratio (SINR) is defined as
\begin{flalign}
    \gamma_n = \frac{p_n|h_n|^2}{\sum_{k=1,k\neq n}^{N}|h_k|^2p_k+\sigma^2},
\end{flalign}
the objective of this work is to estimate the interference power such that the transmitter can effectively adapt its rate to minimize the unexpected interference power variations that affect the received signal quality and SINR. %, deteriorating the system performance. 
More specifically, in URLLC systems  the amount of resources needed must be allocated depending on the outage probability requirement. URLLC packets usually transmit control signals or other application messages where the packet size is small. Due to the small packet size, conventional Shannon-rate formula is not applicable. To address this, Polyanski et al. analyzed the achievable rate and the decoding error probability for finite block-length channels in \cite{Poly}. The channel usage for user $n$ in the finite block-length regime is approximated as
\begin{equation}
    \label{eqn:3}
    R_n \approx \frac{D}{\text{C}(\hat{\gamma}_n)}+\frac{Q^{-1}(\epsilon)^2\text{V}(\hat{\gamma}_n)}{2\text{C}(\hat{\gamma}_n)^2}\left[1+\sqrt{1+\frac{4D\text{C}(\hat{\gamma}_n)}{Q^{-1}(\epsilon)^2\text{V}(\hat{\gamma}_n)}} \right],
\end{equation}
where
\begin{flalign}
    \hat{\gamma}_n=\frac{p_n|h_n|^2}{\hat{\text{I}}_n+\sigma^2},
\end{flalign}
is the predicted SINR, $\hat{\text{I}}_n$ is the predicted interference, $D$ is the number of information bits with decoding probability $\epsilon$, $\text{C}(\gamma)=\log_2(1+\gamma)$ is the Shannon capacity of AWGN channels under infinite blocklength regime, $Q^{-1}(\cdot)$ is the inverse of the Q-function and $V(\gamma)=\frac{1}{\ln{(2)^2}}(1-\frac{1}{\left(1+\gamma\right)^2})$ is the channel dispersion.

A model-based approach for interference prediction was recently presented by Mahmood et al. in \cite{Nurul}. The interference distribution was modeled using a DTMC and predictions were made in such a way that predicted interference is greater than the actual interference with probability greater than or equal to the confidence level $\eta$; resources are calculated using (\ref{eqn:3}). The performance results presented in \cite{Nurul} will be used as performance benchmarks for the proposed NARNN and this approach will be referred to as statistics-based prediction since it considers the entire distribution of interference values; similar assumptions to this statistics-based prediction are considered in this paper: the desired channel has a mean SNR $\gamma_{D}$ and the mean interference to noise ratio (INR) of each interfering link is  uniformly distributed in the range $[\gamma_{I,\text{min}},\gamma_{I,\text{max}}]$,
the desired transmitter transmits packets of $D$ bits with a target block error rate (BLER) $\epsilon$.
% To meet the BLER requirement, the transmitter estimates the SINR and allocates the required resources accordingly, the amount of resources needed are calculated using \ref{eqn:3}. The estimates are obtained based on modeling the discretized interferers' power as a discrete-time Markov chain (DTMC). Hence, the interference values are estimated by using the transition probabilities to predict the next state, i.e., next time slot interference power, given a confidence level. 

% Once and estimate has been obtained, the amount of resources needed have to be allocated depending on the outage probability requirement. 
Different from the aforementioned paper, we propose to employ a novel method to precisely estimate the interference power based on NARNNs. To do so, we first review fundamental concepts of NARNNs in the next section.

\section{NAR Model} \label{sec:narModel}
Interference power in a time series is difficult to forecast accurately due to the random nature of the wireless channel. A linear model would not be able to model the random variations in the wireless channel. For this reason, a NARNN has been used for effective time series prediction of interference. The NARNN is a recurrent dynamic network with feedback connections enclosing layers of the network; thus, the current output depends on the values of the past output \cite{7060655}.  The NARNN can be defined as \cite{NYANTEH2013201} 
\begin{equation}
\label{eqn:2}
y(t)=f(y(t-1),y(t-2),y(t-3),...y(t-n))+ \epsilon (t),
\end{equation}
where $y$ is the data series for the combined interference values of N-1 interferers, $f(\cdot)$ is a transfer function that can be approximated by the neural network, $n$ is the input delay of interference time series and $\epsilon$ is the approximation error.

\begin{figure}[htbp]
\centering
%\includesvg[width=0.5 \textwidth ]{narnnArchUlt.svg}
\includegraphics[width=0.5 \textwidth ]{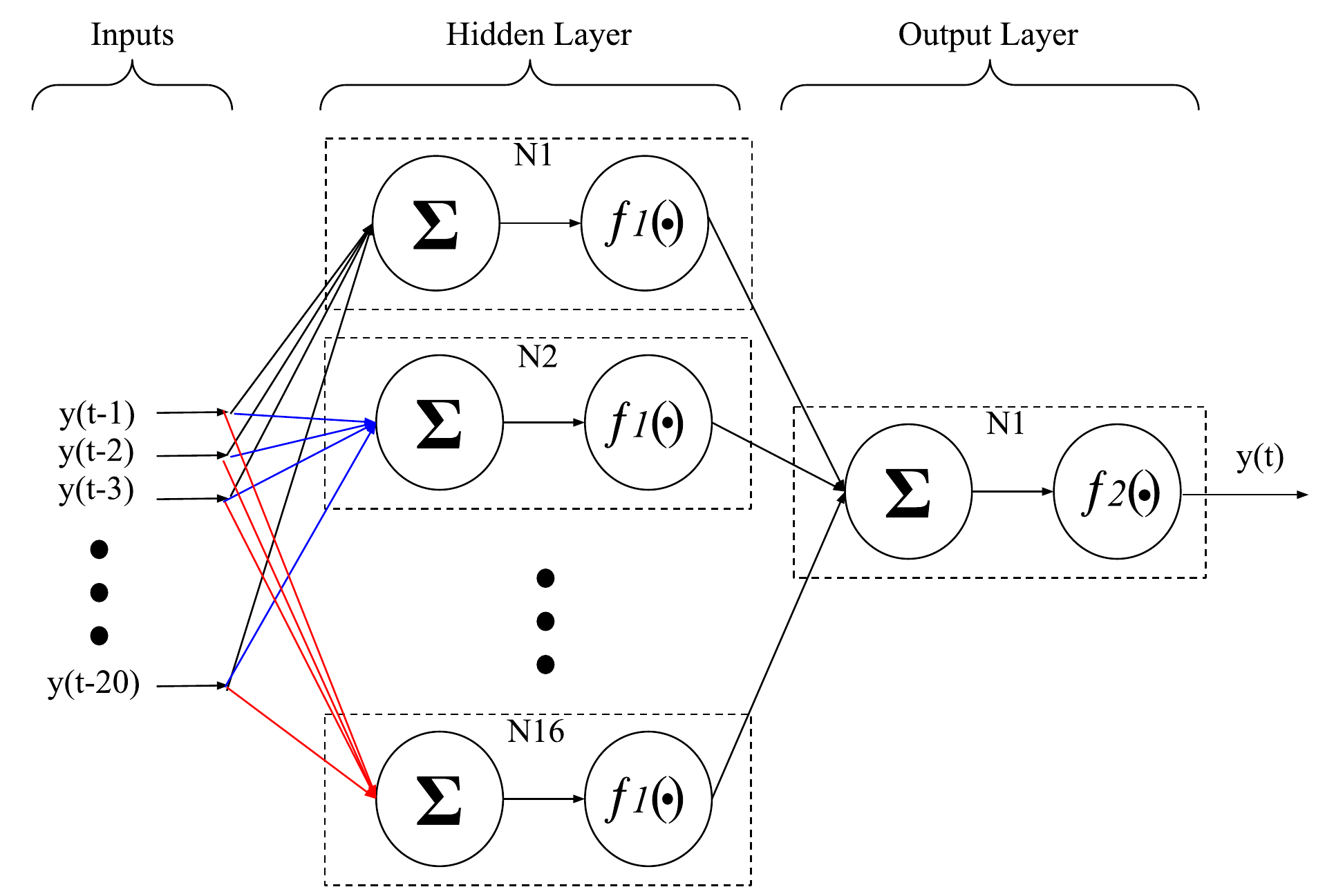}
%\includegraphics[width=0.6 \textwidth ]{Paper.svg}

 %where an .eps filename suffix will be assumed under latex, 
% and a .pdf suffix will be assumed for pdflatex; or what has been declared
% via \DeclareGraphicsExtensions.
\caption{NARNN architecture.}
\label{NNarch}
\end{figure}

%The purpose of training the NARNN is to determine the weights and bias of %neurons by minimizing the error of NARNN output values when compared with the %true values. 
Fig. \ref{NNarch} shows the architecture of the NARNN. It is composed of three layers: input, hidden and output layers. % It is a feedforward neural network 
The NARNN architecture requires specific number of delays, hidden nodes, activation functions and a efficient training algorithm \cite{water}. All these parameters can be optimized by trial and error \cite{en9090684}. We found 20 delays to be accurate enough for prediction of Rayleigh channel interference values. In addition, 16 neurons for hidden layer and both \emph{logsigmoid} ($f_{1}$) and linear activation ($f_{2}$) functions were used in hidden and output layers, respectively.  Levenberg-Marquadt back propagation  (LMBP) was chosen as the training algorithm because of its fast convergence speed and accuracy \cite{trainingAlg,algoMath}.
%\begin{figure}[h]
%\centering
%\includegraphics[width=2.5in]{NNarch.jpg}

 %where an .eps filename suffix will be assumed under latex, 
% and a .pdf suffix will be assumed for pdflatex; or what has been declared
% via \DeclareGraphicsExtensions.
%\caption{NAR neural network.}
%\label{fig_sim}
%\end{figure}

\begin{figure}[htbp]
\centering
%\includesvg[width=0.5 \textwidth ]{narFlow.svg}
\includegraphics[width=0.5 \textwidth ]{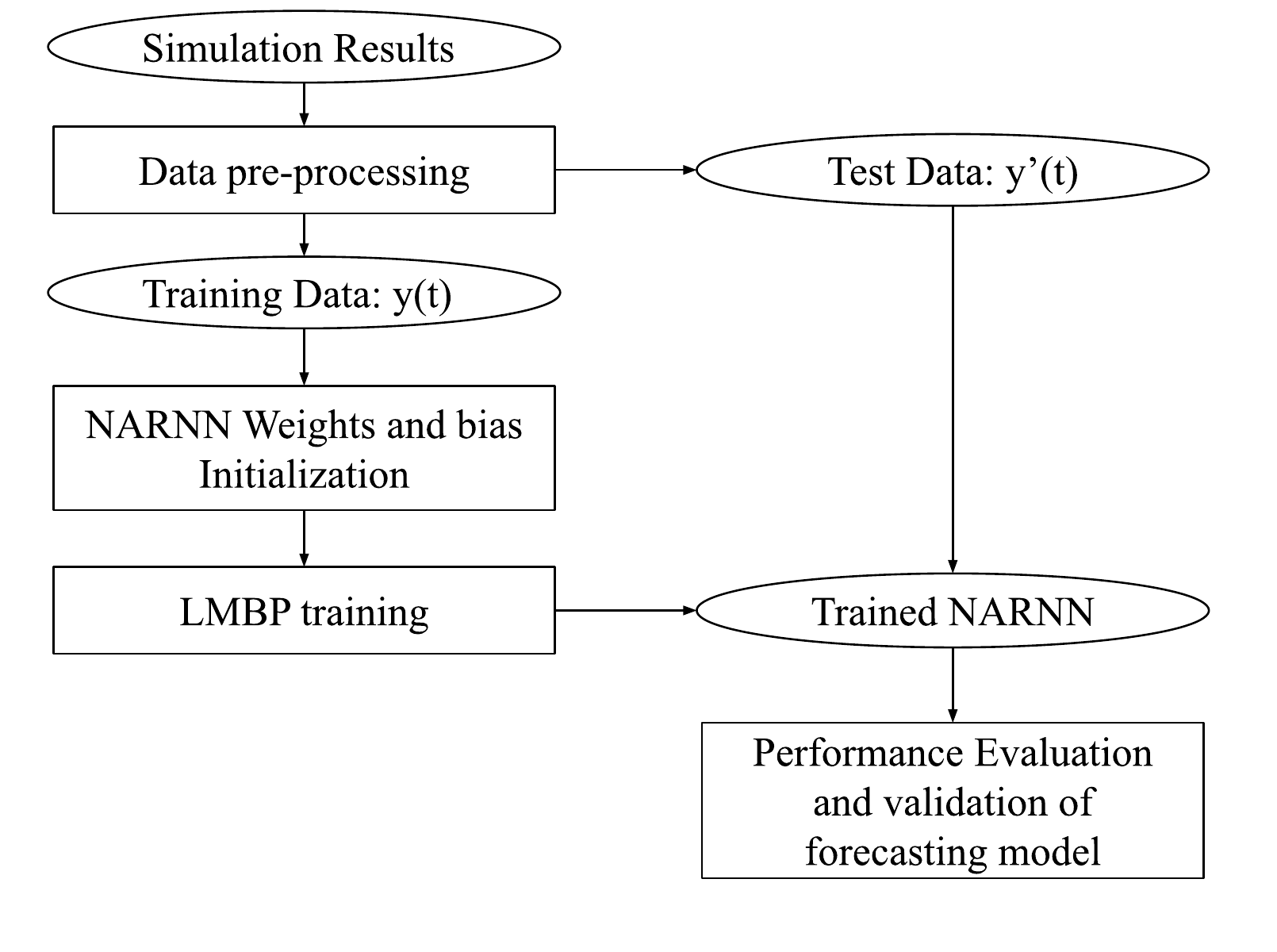}
\caption{NARNN training procedure.}
\label{narFlow}
\end{figure}

The necessary steps for implementing the NARNN are shown in Fig. \ref{narFlow}. First, the time series data sets obtained in simulation was divided in two different sets: training and testing sets, this is the data pre-processing. NARNN weights and bias are later initialized with random values. LMBP was used to adjust their values in every iteration (epoch); as many iterations as needed should be performed to achieve a target error. The final step consists in using the test data to evaluate the performance and forecasting ability of the NARNN. Details about the performance of the NARNN will be discussed in the following sections. The open loop architecture (no feedback loop) was chosen over the closed-loop architecture because predictions were made one-step-ahead. Closed loop architecture is typically used for multistep-ahead predictions~\cite{wind}. For this proposal, the NARNN training data sets consist of interference power time series obtained from computer simulations of downlink transmission with N-1 interferers, all of which are Rayleigh faded.  Details on performance evaluation of the NARNN and resource allocation are discussed in the next section.

\section{Results and Discussion}
\label{sec:results}
% For comparison purposes, the channel data was generated by simulating the downlink of a wireless network with N interferers that do not cooperate among them. In addition a Rayleigh block-fading channel model was used as depicted by Mahmood et al. in \cite{Nurul}.

This AI-based approach has two objectives. The first one is to predict the interference values using the developed NARNN; performance of the NARNN is evaluated by means of MAPE and MSE metrics. The second objective is to allocate resources depending on the NARNN output; the performance is evaluated again after allocating resources by studying the resource usage and outage probabilities achieved by the NAR-based approach and comparing it to the statistics-based prediction in \cite{Nurul} where interference power value is predicted using a discrete time Markov chain (DTMC) and a configurable confidence level. With respect to resource allocation, the predicted interference power needs to be over-estimated to allow the system to meet target outage probabilities ranging from $10^{-1}$ to $10^{-5}$. This over-estimation can be accomplished by using an adjustable factor $1<\alpha<2$ to scale the NARNN output. We denominate this procedure as the Resource Control stage. 

\begin{figure}[htbp]
\centering
%\includesvg[width=0.5 \textwidth ]{stagesNAR2.svg}
\includegraphics[width=0.5 \textwidth ]{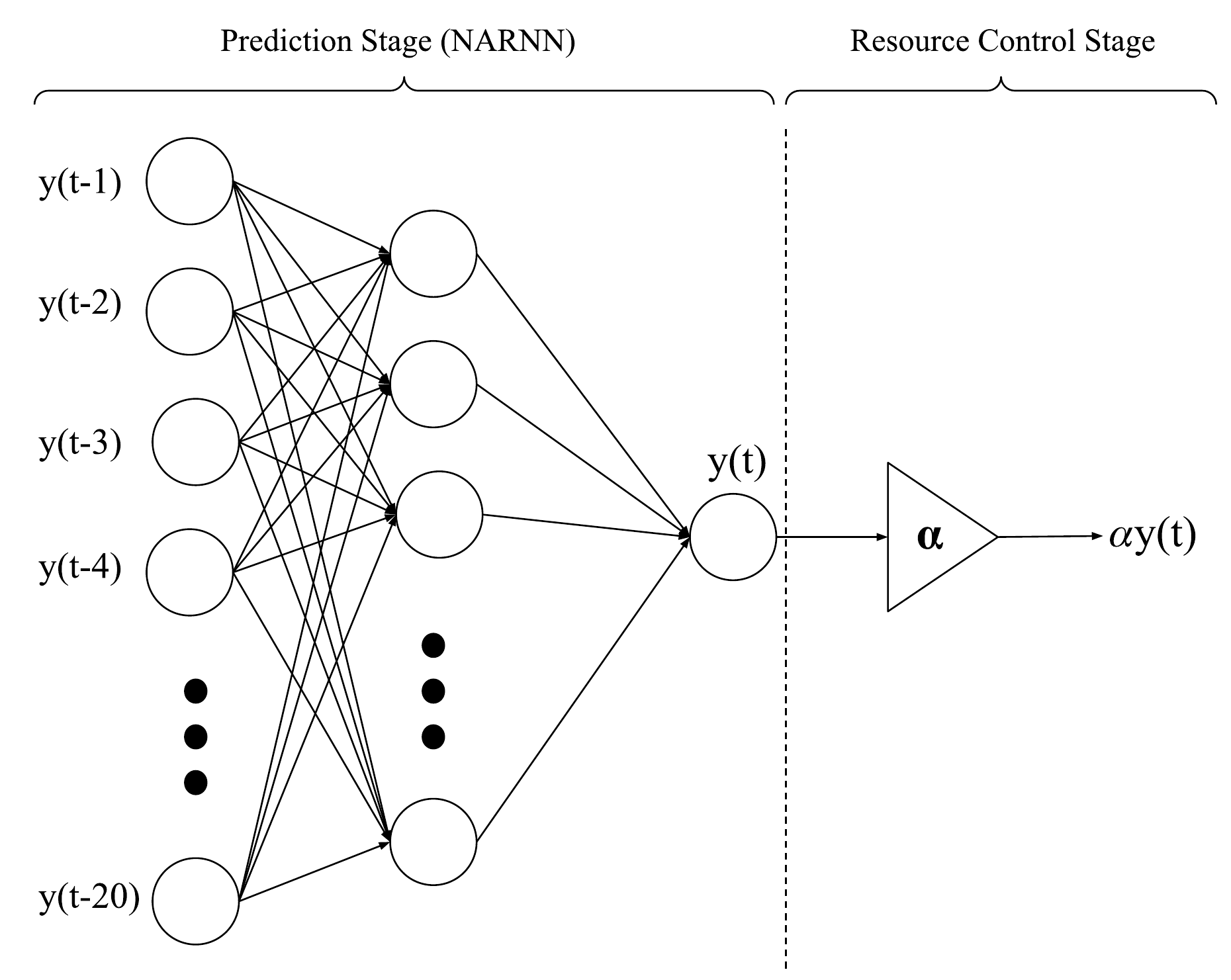}
%\includegraphics[width=0.6 \textwidth ]{Paper.svg}

 %where an .eps filename suffix will be assumed under latex, 
% and a .pdf suffix will be assumed for pdflatex; or what has been declared
% via \DeclareGraphicsExtensions.
\caption{Resource Management Model Proposal.}
\label{solution}
\end{figure}

Fig. \ref{solution} shows the 2-stage process of the proposed NAR-based RRM technique. The NARNN is used in the first step for interference forecasting (namely prediction stage) and the adjustable factor $\alpha$ is used in the second step for regulating resource utilization.

% \begin{figure}
%     \centering
%     \includegraphics[trim = 4.5cm 4.5cm 4.5cm 4.5cm,scale=0.55]{figs/fig1.pdf}
%     \caption{The system model.}
%     \label{fig:1}
% \end{figure}

Regarding prediction stage, the NARNN was trained using  data sets that consist of interference power time series obtained from computer simulations of a downlink with N-1 interferers operating in a Rayleigh channel. The prediction performance of the NARNN was evaluated using  mean squared error (MSE) and mean absolute percentage error (MAPE). Mean square error is given by
\begin{equation} 
\label{eqn:6}
 \text{MSE} = \frac{1}{n} \sum_{i=1}^{n} (y_{i}-\hat{y}_{i})^2,
\end{equation}

MAPE makes comparisons easier because it is percentage-based \cite{mape}. Mean absolute percentage error is defined as
\begin{equation} 
\label{eqn:7}
 \text{MAPE} = \frac{1}{n} \sum_{i=1}^{n} \frac{|y_{i}-\hat{y_{i}}|}{y_{i}} \times 100,
\end{equation}

\begin{figure}[htbp]
\centering
%\includesvg[width=0.5 \textwidth ]{trueAndNN2.svg}
\includegraphics[width=0.5 \textwidth ]{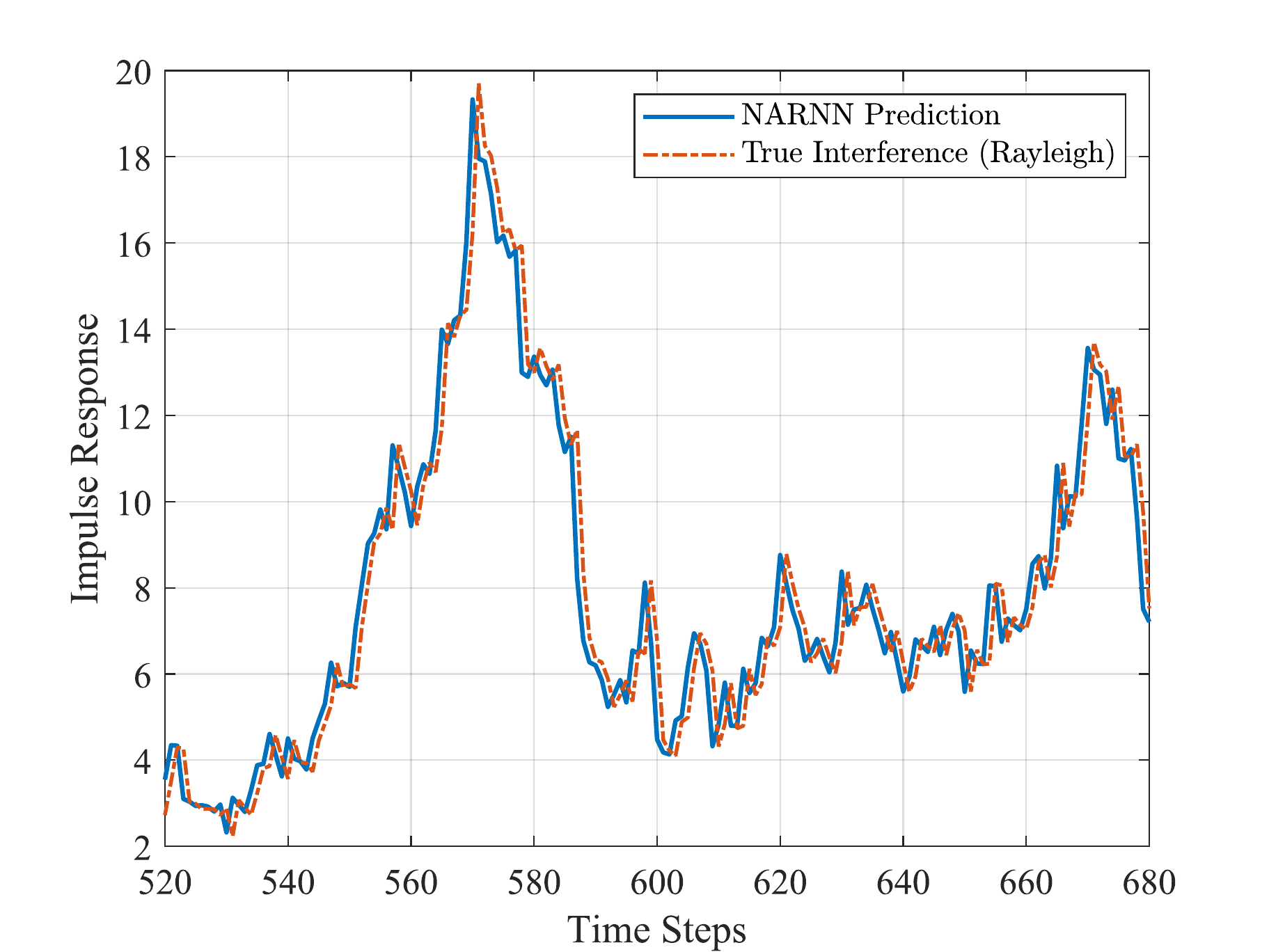}
%\includegraphics[width=0.6 \textwidth ]{Paper.svg}

 %where an .eps filename suffix will be assumed under latex, 
% and a .pdf suffix will be assumed for pdflatex; or what has been declared
% via \DeclareGraphicsExtensions.
\caption{NARNN predictions for a fixed range of time steps.}
\label{NNpred}
\end{figure}

The variables $y_{i}$ and $\hat{y_{i}}$ represent real and predicted interference values respectively. Impulse response corresponding to the NARNN in Prediction Stage is shown in Fig. \ref{NNpred}. The best-performing NARNN was used to forecast interference power. For comparison purposes, the true values from simulations have also been plotted. The NARNN achieved a promising MSE of 0.5447 and MAPE of 7.816\% in the optimal configuration. 
% 16 neurons in the hidden layer, 20 delays and log-sigmoid and linear activation functions in the hidden and output layers respectively.
% These parameters were found by trial and error. 
To find the optimal number of neurons tests were made with both \emph{log-sigmoid} (namely logsig) and \emph{tangent hyperbolic sigmoid} (namely tansig) functions. Table \ref{tab1} shows the results of the test with a variable  number of neurons with \emph{logsig} and \emph{tansig} functions as activation functions in the hidden layer. These results show that there is not a significant benefit of increasing the number of neurons. \emph{Logsig} performed at its best  using 16 neurons with an MSE of 0.5447 while \emph{tansig} reached peak performance with 14 neurons and an MSE of 0.5446. While there is a slight improvement with the \emph{tansig} function with 14 neurons, we observed a higher number of epochs needed for training when compared to the \emph{logsig} function. \emph{Tansig} function required 64 epochs while \emph{logsig} was trained with 19 epochs. It was concluded from these experiments that the developed NARNN required 16 neurons for the best overall performance.

\begin{table}[htbp]
\caption{MSE and MAPE for different Number of Neurons}
\begin{center}
{\scalebox{0.95}{
\begin{tabular}{|c|c|c|c|c|c|c|}
\hline
\textbf{Activat.}& \textbf{Metric} &\multicolumn{5}{|c|}{\textbf{Number of Neurons}} \\
\cline{3-7} 
\textbf{Funct.} &  & \textbf{\textit{8}}& \textbf{\textit{12}}& \textbf{\textit{14}}& \textbf{\textit{16}}& \textbf{\textit{18}}  \\
\hline
Logsig & MSE & 0.5449 & 0.5458 & 0.5455  & 0.5447 & 0.545 \\

\cline{2-7}
 & MAPE & 7.828& 7.836 &  7.832 & 7.816 & 7.829 \\
\hline
Tansig & MSE & 0.5452& 0.5449 & 0.5446  & 0.5454 &0.5450\\
\cline{2-7}
 & MAPE & 7.830& 7.824 & 7.831 &7.825 & 7.826 \\
\hline

\end{tabular}}}
\label{tab1}
\end{center}
\end{table}

 To study the effect of increasing the number of delays in the performance, multiple NARNN versions were trained with a different number of delay taps, specifically: 2, 20 and 50 delay taps. MSE was almost identical for these 3: 0.546, 0.5447 and 0.545 respectively with MAPEs around 7.8\%. These results demonstrate that the NARNN is capable of modelling the Rayleigh channel with a small number of neurons and delays and that gains in performance are almost nonexistent when tuning these parameters. Given these results, 20 delays and 16 neurons were chosen to get a marginal improvement while keeping the complexity of the NARNN low.
 
With respect to the Resource Control Stage, $\alpha$ can be adjusted once a sufficient forecasting performance has been attained at the prediction stage, it is important to note that regulating $\alpha$ leads to changes in resource usage and outage probability. Since Resource Control Stage consists on a factor multiplication, it is not possible to evaluate its performance, we rather exhibit the performance of the entire NAR-based approach in two ways. The first one, consists in adjusting the $\alpha$ factor to get the NARNN resource usage to equal the statistics-based estimate resource usage; in this scenario, the objective is to study the mean outage probability of these two techniques having the resource usage fixed. The second experiment is similar but instead of targeting the resource usage with $\alpha$, we aim at the mean outage probability. The factor $\alpha$ is tuned to lower the resource usage of the NARNN while matching the statistics-based estimate mean outage probability. The objective is analogous to the first experiment, to study the resource usage of both techniques with a fixed mean outage probability.

\begin{figure}[htbp]
\centering
%\includesvg[width=0.5 \textwidth ]{outRes.svg}
\includegraphics[width=0.5 \textwidth ]{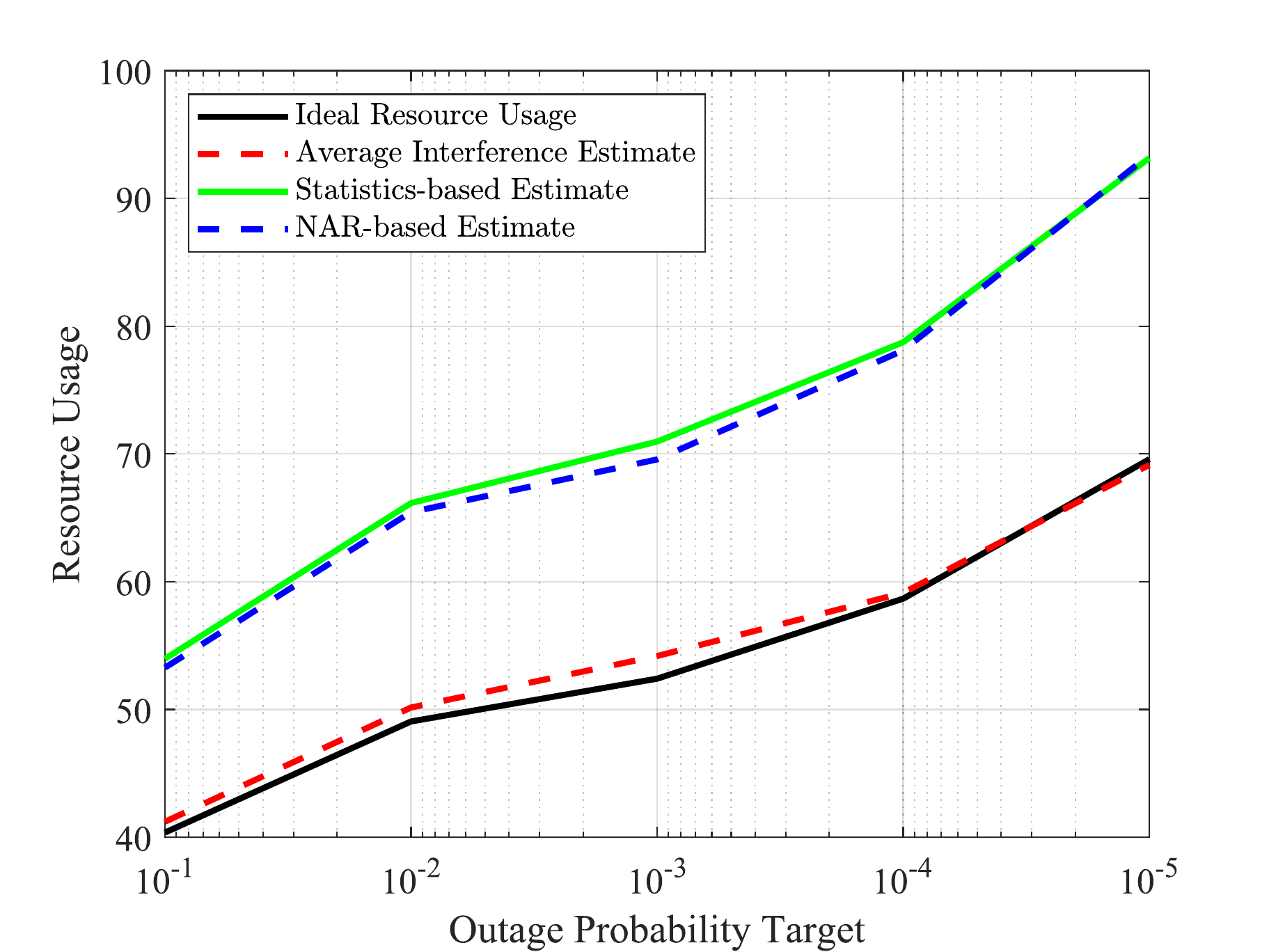}
%\includegraphics[width=0.6 \textwidth ]{Paper.svg}

 %where an .eps filename suffix will be assumed under latex, 
% and a .pdf suffix will be assumed for pdflatex; or what has been declared
% via \DeclareGraphicsExtensions.
\caption{Resource Usage of NAR-based estimate with $\alpha=1.45$.}
\label{1out}
\end{figure}

 Fig. \ref{1out} shows the ideal, statistics-based and the  NAR-based prediction resource usage  for outage probabilities ranging from $10^{-1}$ to $10^{-5}$. It was found empirically that a value of $\alpha=1.45$ equals the NARNN resource usage with the statistics-based prediction resource usage. Note that ideal resource usage has been calculated using a genie-aided estimator and is considered to be the optimum performance bound while the average interference is calculated using an IIR filter \cite{Nurul}. Also recall that the $\alpha$ factor may be regulated dinamically to meet target outage probabilities of systems that differ from our proposal; increasing $\alpha$ will lead to smaller outage probabilities at the cost of higher resource usage.

\begin{figure}[htbp]
\centering
%\includesvg[width=0.5 \textwidth ]{outMean2.svg}
\includegraphics[width=0.5 \textwidth ]{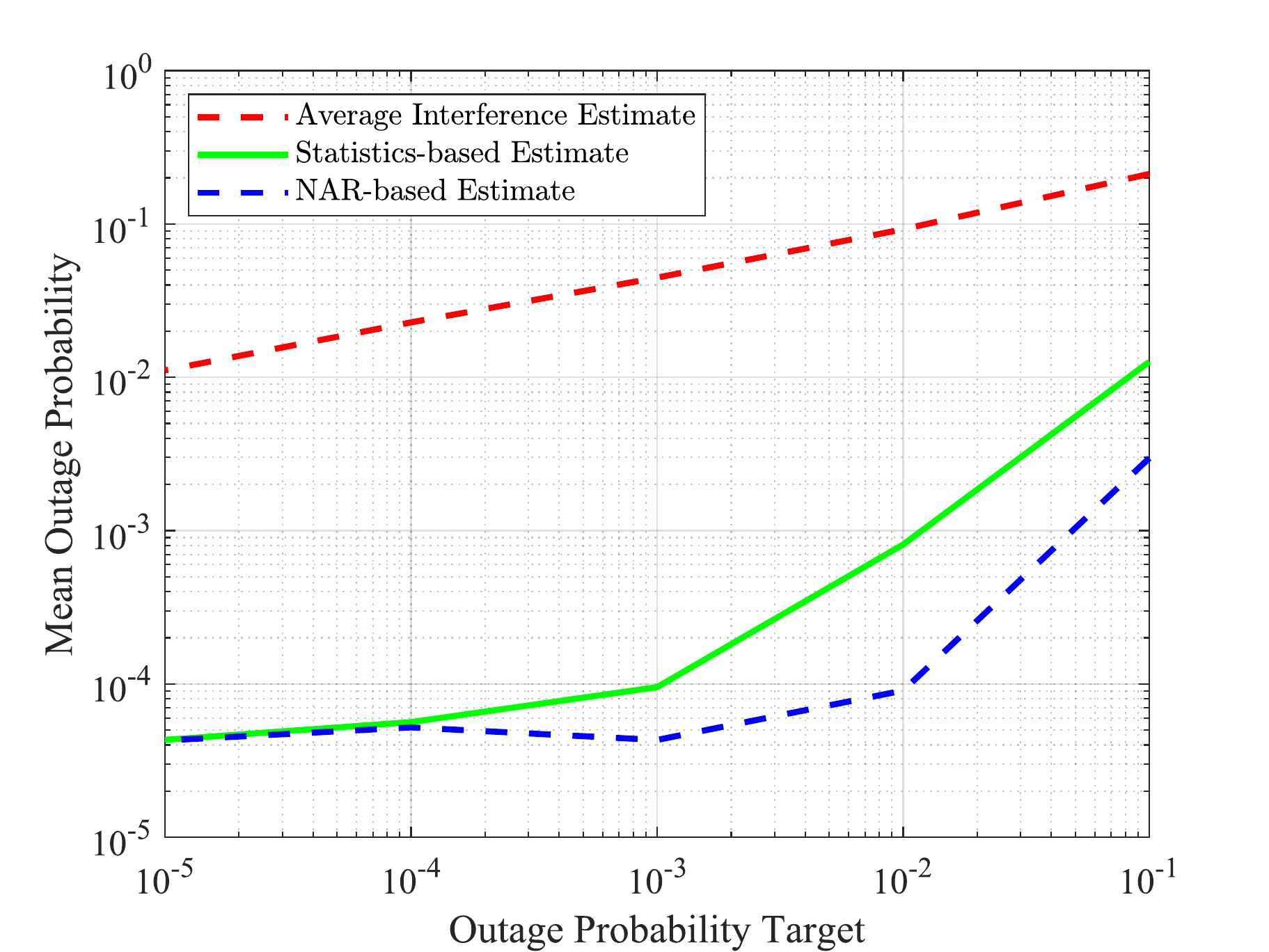}
\caption{Outage probability of NAR-based estimate with $\alpha=1.45$.}
\label{outRes}
\end{figure}

\begin{figure}[htbp]
\centering
%\includesvg[width=0.5 \textwidth ]{RU120.svg}
\includegraphics[width=0.5 \textwidth ]{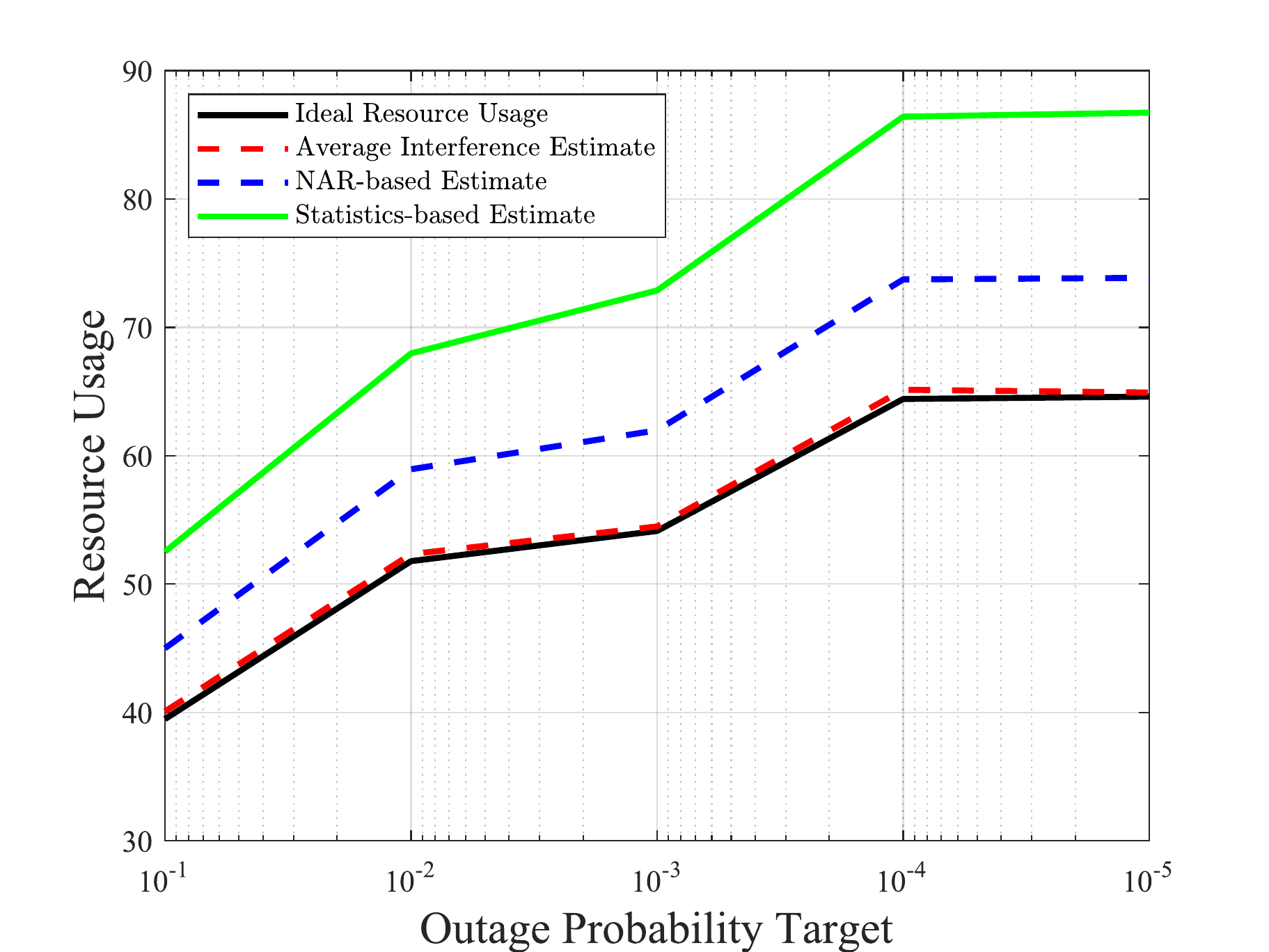}
\caption{Resource Usage of NAR-based estimate with $\alpha=1.2$.}
\label{RU120}
\end{figure}

\begin{figure}[htbp]
\centering
%\includepdf[width=0.5 \textwidth ]{8.pdf}
\includegraphics[width=0.5 \textwidth ]{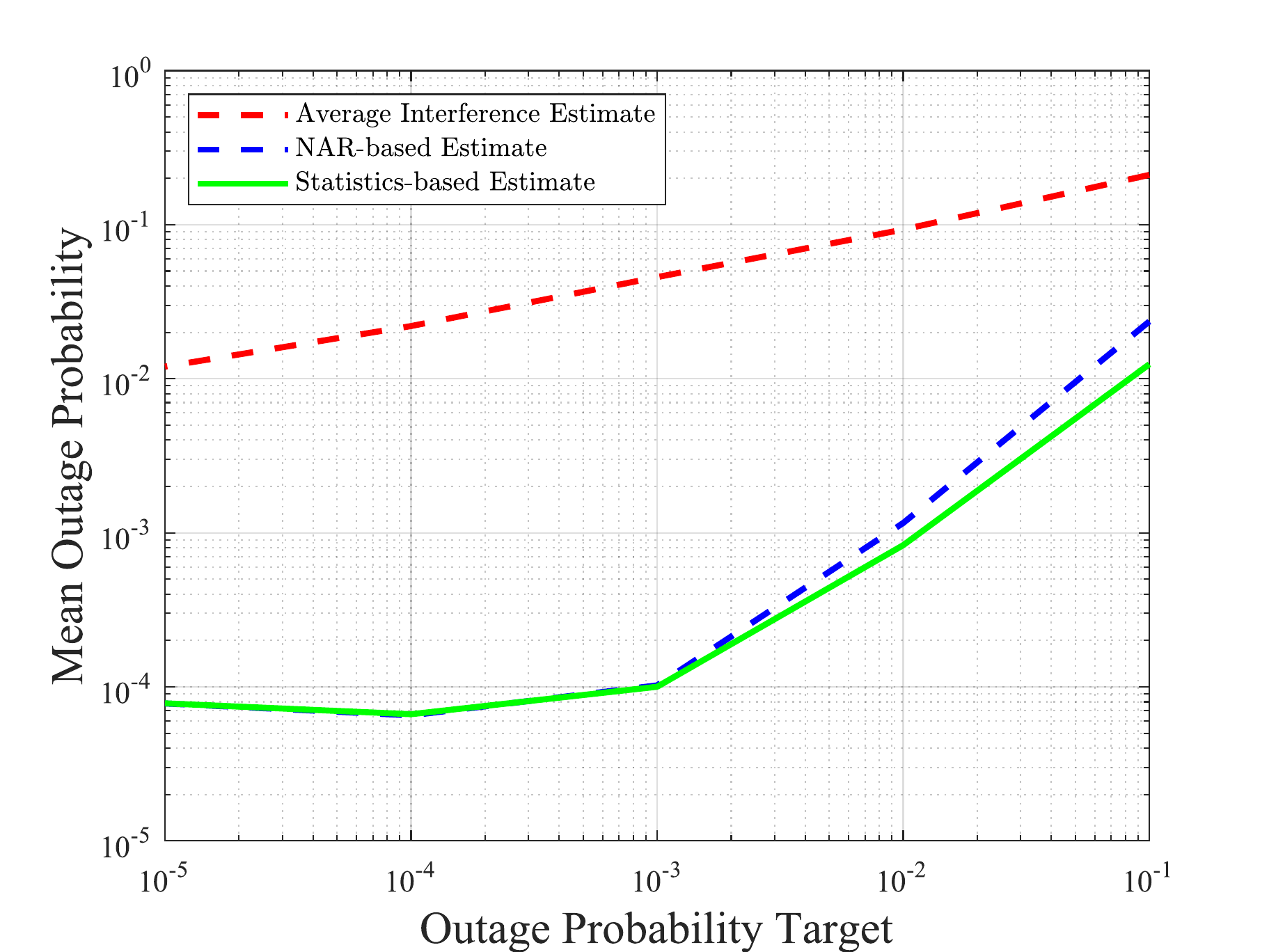}

 %where an .eps filename suffix will be assumed under latex, 
% and a .pdf suffix will be assumed for pdflatex; or what has been declared
% via \DeclareGraphicsExtensions.
\caption{Outage probability of NAR-based estimate with $\alpha=1.2$.}
\label{OP120}
\end{figure}

 The mean outage probability of the NAR and statistics-based approaches for $\alpha=1.45$ and the same range of target probabilities ($10^{-1}$ to $10^{-5}$) has been plotted in Fig. \ref{outRes}. This plot shows that the NAR-based approach is capable of reducing the mean outage probability considerably by using the same amount of resources as the statistical approach. There is a reduction in outage probability of factor up to 1/10 when using the NARNN with  $\alpha=1.45$. 
It is possible to target the outage probability of the statistical approach instead of the resource usage by choosing $\alpha=1.2$. Fig. \ref{OP120} shows mean outage probability for this $\alpha$ value. Both NAR and statistics-based estimates have similar outage probabilities but the resource usage of the NARNN went down by approximately 15\% compared to the statistics-based estimate as seen on Fig. \ref{RU120}. 
% In other words, the same performance can be achieved using less resources if the NAR-based approach is in use. 
%  Fig. \ref{RU120} shows the actual benefit of the NAR-based prediction comparing the outage probabilities of the statistical and NAR-based approaches. It can be seen from the graph that NAR-based approach outperforms the statistical approach. These results demonstrate that it is possible to outperform the statistical approach by using the NARNN and at the same time it is possible to control the resource usage using the factor $\alpha$.

% Regarding the coefficient gain placed at the output of the NARNN, resource usage and mean outage probability has been plotted for several C values to study its behaviour. It can be seen that a higher C value provides higher resource usage while having less mean outage probability. This behavior is to be expected, although some improvements were achieved in contrast with Nurul's algorithm. We needed approximately 14$\%$ less resources to get a similar mean outage probability. At the same time we achieved a remarkable reduction in outage probability while using the same amount of resources as the original algorithm. Figure [citation figure] and Figure [citation figure] show resource usage and mean outage probability for different values of coefficient C respectively.

\section{Conclusion}
\label{sec:conclusion}
Having knowledge of the channel is of major importance to avoid high error rates and allocate resources accordingly. In addition, wireless channels behaviour is chaotic which makes prediction approaches very appealing. The focus of this proposal are URLLC scenarios, for this reason, predicting the channel behaviour is not the only objective, meeting URLLC target outage probabilities is also a major goal. To target these two objectives, we have proposed the use of a NAR-based RRM technique with two stages: interference prediction and resource control stages. For interference prediction, a NARNN was used due to its promising prediction capabilities. For resource control, a factor $1<\alpha<2$ is used to scale the output of the NARNN and to regulate resource usage. 
% The wireless system considered consists in a downlink with N interferers and Rayleigh fading.
The NARNN provided very accurate predictions showing a promising MSE of 0.5447 and a MAPE of 7.816\%. It was found by trial and error that the optimal configuration for the NARNN was 16 neurons in hidden layer, 20 delays and a log-sigmoid activation function. The performance of the whole system (NARNN prediction and resource control) was compared to an statistics-based approach. For $\alpha=1.2$, the NAR-based prediction matched the outage probability of the Statistics-based approach while using up to 15$\%$ less resources. In addition, the outage probability is improved by an order of 10 times when using the NARNN with $\alpha=1.45$.
%  For $\alpha=1.45$  the NAR-based prediction matched the resource usage but outperformed the statistics-based based approach by reducing the outage probability in factor of up to 10 times. 
In summary, the NARNN showed promising predictions of interference and allowed a more efficient resource allocation when compared to statistics-based estimates. In this study NARNNs has been used to predict interference in a Rayleigh channel, however, we plan to consider a more complex fading model in our future works.


% \section*{Acknowledgment}

% The authors would like to thank CWC Professors for providing unique opportunities for research AI applications in Wireless Communications.




\bibliographystyle{IEEEtran}
\bibliography{IEEEabrv,biblio}
%\begin{thebibliography}{1}

%H.~Kopka and P.~W. Daly, \emph{A Guide to \LaTeX}, 3rd~ed.\hskip 1em %plus

%\end{thebibliography}

\end{document}